# Modeling and Self-Configuring SaaS Application

**Nadir K.Salih, Tianyi Zang**

School of Computer Science and Engineering, Harbin Institute of Technology, Harbin, Heilongjiang, China

**Abstract -** *The main objectives of SaaS application are to make the management and control of software easier and take the management strain away from consumers. However, it also leads to software services available globally and this has been realized in our paper by designing a new model for SaaS application. The three levels we have classified in our model easy adapted to workflow and services. From the application layers meat-model description we discovered a new algorithm for the self-configuration of SaaS application. We used a feature model to define the variation of our model's management levels. The Xml file obtained from the feature model gave interactive communication between three levels and our new self-configuration algorithm. That increased the performance by selecting from the web a suitable configuration for every level. We have explained all the processes by an online booking example. Finally we present a conclusion and future work.*

## 1   Introduction

Modeling SaaS application is very important field and building a SaaS by leveraging existing technology is a challenging issue and needs brand new software technology [1]. It is useful for both business and educational purposes, such as businesses can be easily adopted in several domains, like healthcare, education and OA (Office Automation) for this to be modeled, the SaaS application [2] [3] demands new requirements. In this paper we have drawn a new model [4] [5] of SaaS application.   We have summarized our contributions as follows:

- Built new model for SaaS application.
- By meta-model defined four layers to compose the system and showed the associations and dependencies of the layer elements.
- Demonstrated the relationship between the three levels in our model by a workflow as a business process layer
- We observed the necessity of sharing the workflow (can share other things, e.g. software components, SLA/QoS, etc) in each level and how it can improve efficiency and better control customer service.
- We have classified services of SaaS application according to three levels. Some services are done by the user; others are by the tenant and some by the provider.
- Increased the quality of system by showing it has different levels of services which can serve by order of

importance.  The service of the provider it is more important than the service of the tenant and tenant services are more important than user services.
- Self-configuration of the algorithm to dynamically configure SaaS components.
- Commonality and variability are indicators for components costs.

We organized this paper by beginning with the design of the new general architecture for SaaS application in section 2. Depending on the model driven development we derive SaaS meta-model layers in section 3. That classifies the SaaS application management in three levels. To demonstrate this new opinion we take online booking SaaS application as running example in section 4.  In section 5 we have described the service architecture for SaaS application. We realized self-configuration of the model by a new algorithm in section 6. Section 7 described the related work. Finally, we present the conclusion and point to future work.

## 2   Architecture of SaaS Application

System modeling is a very important issue in software engineering, because it has great importance in system development. Thus, we have defined our architecture of SaaS application, and described our model by using the meta-model concept to show we could easy achieve management by the new model. Application architecture specifies that technologies are to implement one or more information systems in terms of data, process, and interface, and that these components interact across a network [6]. Architecture is a transferable abstraction of a system [7]. As we study from recent researches architecture development of SaaS is a large part of the application. Our novelty here is to create a conceptual model for SaaS application as depicted in figure 1.

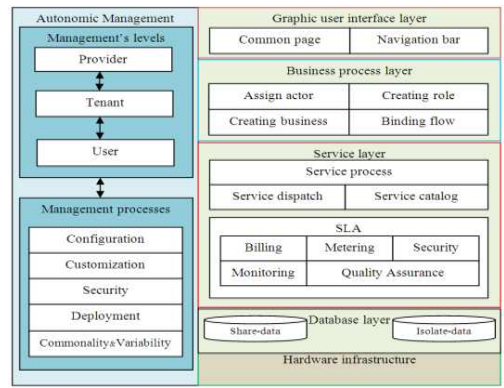

Fig 1 Architecture of SaaS Application

This architecture includes main three parts:

• Application layers contain four layers beginning from the graphic user interface (GI) that uses the web page and navigation bar to communicate to the user and SaaS application. The second layer is the business process (BP) to show the workflow for business by defining some roles and actors activities. In the service layer service (S) process is determined by the dispatch manner and catalog. In addition the service level agreement has been defined for some services like billing, monitoring, QA, metering, and security. The final layer is the database (DB) layer which shares the common data and isolates variable data.

• The hardware infrastructure includes all hardware resources working in SaaS application servers, storages, network, etc. The allocation and placement algorithm is used to optimize these resources.

• The autonomic management manages all management in SaaS application and will be self-managed in the three levels of provider, tenant and user. This will be applied in the application layers to manage the processes (configuration, customization, security, validation, commonality, and variability).
Adapting the same application in the case of multiple users to somewhat different and specific needs of a certain user is important therefore creating a new architecture suitable for development is needed. The new concept in our proposed architecture of SaaS application is the base in three levels adapting to develop SaaS by adjusting to the tenant's instant functions from the provider level. We have looked to adaptive to different instances for all users from the tenant level. Also, adjusting user requirements from the user level are controlled.
*The goals of modeling are:*

• Develop architecture for SaaS approach based on three levels to realize organization and user requirements.

• Configuration and adaptation of SaaS applications must be performed.

• Customized adaptation for every level to ease management of SaaS application.

## 3 Meta-Model of SaaS Application

Looking at the proposed model for SaaS application three management levels have been classified that are depicted in figure 2. According to the kind of service SaaS system can determine the level of management. The reasonability of this classification is a variation [8] of the application layers from level to level.

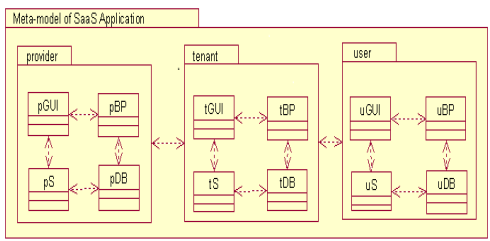

Fig 2 Three Management Levels for SaaS Application

All application layers can be variables in the provider level for different tenant requirements. Likewise, in the tenant level all application layers are changeable for different user requirements. However in the user level we observed it is the same as in GUI, BP, and S, but DB it different from user to user. The general meta-model [9] of these layers are depicted in figure 3 below.

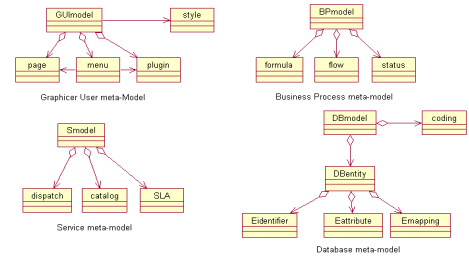

Fig 3 Meta-model of SaaS Layers

## 4 Demonstrate SaaS Meta-Model

As we mention our SaaS model has three levels of management including the provider, tenant and user. Every level has different managements for the application layers, which are defined in the upper meta-model of SaaS layers. We can take an on-line hotel booking example to demonstrate this model as seen in figure 4. The provider is a highly configurable service that travel agencies can use for booking hotels on behalf of their customers. For that we can say the provider is the administrator for all travel agencies. The travel agencies look like tenants and customers are users that want to book a travel service.

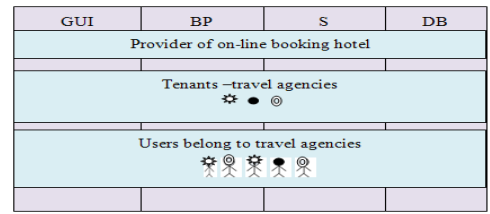

Fig 4 SaaS Application of on-line Booking

Depending on the meta-model layers we describe our own model of SaaS application. At first, the provider is the administrator of all the travel agencies to management the activities that appeared through the layers:

• The Graphic User Interface(GUI) has different style for the GUI layer which has various types like standard menu and tree menu to display page or plug ins as requirement the from tenant.

• In the Business Process (BP) the provider puts the business logic in a formula. It can be a variable from agency to agency. And the workflow defined by a sequence, branch and return according to the agencies requirements. In addition the status is used to show that the software is open or closed in each different status.

• Services (S), can be different for the service dispatcher and service catalog between agencies. For example if the agency categorizes the services as and VIP they will be dispatched differently and then cataloged which means the sending, indexing services from provider to tenant are not the same. However the variation in the service level agreement for many services billing, metric, and security can be according to agency requirements.

• Data Base (DB), in this layer we should define as any agency by a unique identifier. And attributes of the data will normally be different from one agency to another. However, the mappings that describe the relationship between different entities vary in the data.

The second level is the tenant that corresponds to the travel agency, and in our example to the management of all users' activities inside the layers:

• The Graphic User Interface (GUI) is the job of the travel agency to show a suitable style interface for the users as a classification for the user as a normal or VIP user.

• In the Business Process (BP), the business logic can be different from user to user so that travel agency can use a different formula according to the type of user. The workflow can be a variable in this system like low season is different from high season booking. However it defines the status of the system as open or close in various cases.

• In the Service layer the travel agency sending and indexing the services depends on the type of user, and the classification of the service for different costs. This will then be applied according to the service level agreement between the agency and customers.

• In the Data Base the travel agency defines any user by an identifier because it is unique for every user, and attributes data can be different from user to user. In addition, the mapping that describes the relationship between entities will vary.

The third level is user that can communicate with on-line travel agency for hotel booking. In this level the management for SaaS application layers is defined as:

• The variation in graphic user Interface is defined by the tenant or travel agency and it needs management if the user uses a different machine such as the Windows client program running in a PC with the resolution of 1680×1050, a smartphone application with the resolution of 640×480, and a tablet application with the resolution of 1024×768. Moreover, in the business process the logic and workflow is the same put forward by the travel agency. However, in the service layer introductions from travel agency are according to user requirements. While the data base layer needs management because it is different from user to user in the identifier, attributes and mapping relationships for different entities.

In the relationship between the three levels we can consider the workflow in our example of online booking with the hotel as the provider level and is managed at the tenant level or by the travel agencies. Figure 5 shows the process of booking when the request reaches the travel agencies. Then, they can begin to display the information that is filled out by the customer as they have an office to check this data and submit it to be accepted or rejected. This sequence is the same in the two travel agencies, but the second travel agency has a difference in workflow due to the manager check. Here the provider can share a customizable workflow for multiple travel agencies using the assembled workflow.

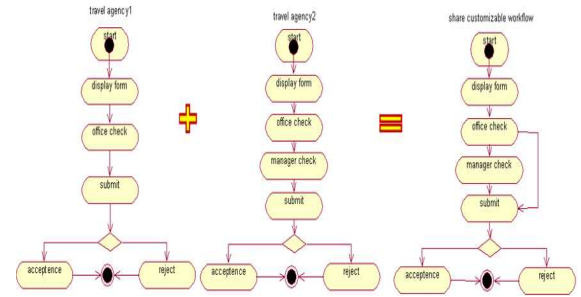

Fig 5 Share Customizable Workflow on the Provider Level

Here the relationship between the provider level and tenant level is a sharing customizable workflow. This can be managed and controlled by many travel agencies in a process by which it shares the same sequences. Our model realized the benefit for a business process by minimizing the many workflow processes in sharing a one workflow process. The relationship between the tenant level and user level can define by the booking process from customer to travel agency. For example, customers in one travel agency web begin by browsing and searching for bookings and payments to finish the transaction. Another travel agency after searching and booking lets the customers to make another search to see new options for booking as depicted in figure 6.

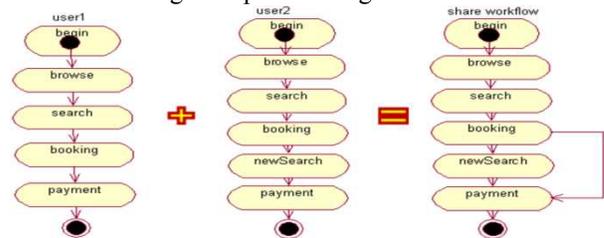

Fig 6 Share Workflow on the Tenant Level

Share workflow can eliminate many unnecessary steps that lead to increased efficiency more than other strategies [10]. It also improved consistency and control result for better customer service. From this relationship between the tenant level and user level we can obviously see our model easily manages and adapts to SaaS application.

## 5    Service Architecture in SaaS Application

To adapt and manage SaaS application we should understand the service architecture represented in our model. Then the feasibility will be clear of our novelty in classifying our model in the three levels of management. The online booking hotel running example will illustrate this principle. Though the web service SaaS application provides different services as defined in our model as a variable from level to level.  From figure 7 we classified our concrete service in $C_{Si}=$ $\{s_{pi}, s_{pi} \ldots s_{pn}, s_{ti}, s_{ti} \ldots s_{tn}, s_{ui}, s_{ui} \ldots s_{un}\}$ , $I \leq i \leq m$, $s_p$, $s_b$, $s_u$ are provider service, tenant service, and user service, respectively. These concrete services obtain the same abstract services as from a functional view, which can be defined by the application layer in formal methods.

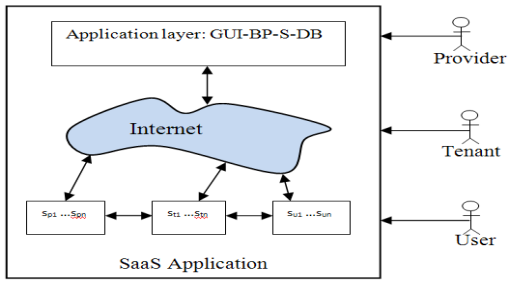

Fig 7 Service Architecture in SaaS Application

To explain this we return to our example of the online hotel booking and see the workflow of this system as appeared in the activity diagram below in figure 8.

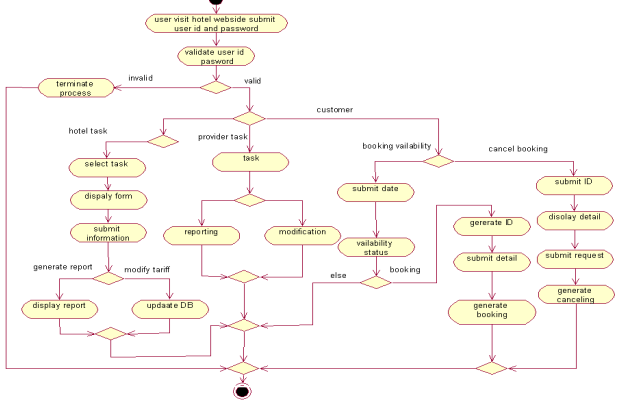

Fig 8 Activity Diagram of Online Hotel Booking SaaS System

Looking at this workflow diagram we start executing as soon as user visits the hotel administrator web side and submits a user ID and password. The SaaS application validates this data and if it is invalid it automatically terminates the process. Here, because the data is valid, it will go on to make the diction from our three levels of management. This user is the customer of the SaaS application and then will provide two services as the different workflow booking are available. The user is the tenant or administrator of the hotel and looks like the travel agencies. This level has services managed by the system display in the form of a customer, submits the information, modifies the tariff, and generates the required report for the customer. The user can be the last level of our model management and is the provider that administrates for all travel agencies here the SaaS application which will provide services to management agencies like reporting for all events and modifying (delete, update, add) the travel agencies data. The SaaS online booking hotel system associates the following abstract services:

- *Available booking*, which lets the customers book a hotel.
- *Canceling booking* prevents customer from booking the hotel.
- *Modify tariff*, the price may change in low season of booking.
- *Generate customer report*; display some reports to the customer.

- *Modify travel agency data*, update, delete and add travel agency data by the provider.
- *Reporting travel agency* displays payment, resources and all management activities of the travel agencies from the provider.

We define two abstract services from each level as follows:

as$_{u1}$ = *Available booking*
as$_{u2}$ = *Canceling booking*
as$_{t3}$ = *Modify tariff*
as$_{t4}$ = *Generate customer report*
as$_{p5}$ = *Modify travel agency data*
as$_{p6}$ = *Reporting travel agency*

## 6  Self-Configuration SaaS Application

By Self-Configuration Algorithm SAAS application (SCAS) and the model driven development approach can be used to implement self-management for SaaS application. The autonomic diagnosis, failures and performance reconfiguration that is required for repair can occur in every layer. The constraint model to check the data conformance has been used in the meta-model to specify constraints. We defined the monitoring model to the instrumentation for collecting data about system behavior. It is very important to reference the architectural entities in reconfiguration or what should happen in any given condition, for what is suitable in the meta-model for monitoring requirements for the environmental and constraint model. We defined the meta-model for the runtime model to reduce the managing complexity during runtime. The prediction method is used to select a suitable configuration [11].To realize self-configuration for SaaS application we should monitor the requirements and environmental conditions. The model systems can be revised and used to generate new codes automatically. The model and meta-model can be control by some constraints. We have used the feature model to define the variation for all meta-model layers in three levels. Simply, we take the provider configuration from the provider to the tenant as an example seen in figure 9. This feature model shows the variations of the configurations for every layer according to constraints that defines the features.

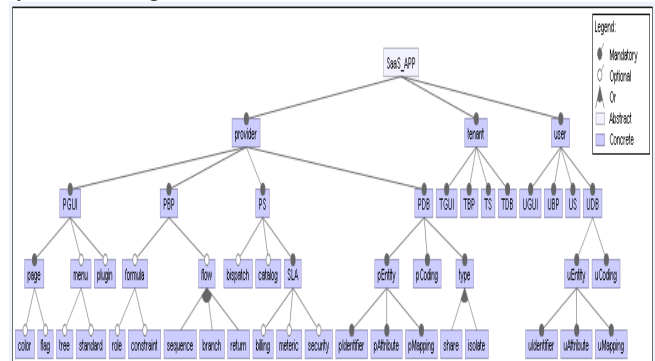

Fig 9 Feature Model for SaaS Meta-model Layers

## 6.1 Logic of SCAS Algorithm

To simplify understand this variation we represent this feature model in a hyper-arc,e, with a multiplicity value, where $mv = [min...max]$, whose tail (feature) is selected, no less than the min and no more than the max features of the hyper-arc's head (child features) should also be presented in the configuration.

$H = (V, E)$, where $V = \{v_1, v_2, v_3, .... , v_n\}$ is the finite set of vertices (or nodes)

$E = \{E_1, E_2, E_3, .... , E_m\}$ / $E_i \subseteq V$ $i = 1,...,m$ is the set of hyper-edges.

$E_i = (T(Ei), H(Ei)) = T(Ei) \subseteq V \wedge H(Ei) \subseteq V$

$E_i$ is a directed hyper-edge , where $T(E_i)$ is the set of tail nodes and $H(E_i)$ is the set of head nodes of $E_i$.

When $|H(e)| = 1$ (children's cardinality set is one):

- If $min = 1 = max$, the feature is mandatory, and should be present if the parent, or it is a required constraint and then the child should also be present [1..1].

- If $min = 0$, $max = 1$, the feature is optional [0..1]

When $|H(e)|>1$(children's cardinality set is more than one):

- if $min = 1 = max$, it is a XOR alternative feature group, and only one child should be present at most if the parent is present.

- if $min = 0$, $max = 1$, it is an optional feature group, and the child features can be present or not as long as its parent is present, or it is a mutex constraint and at most one of the child's features can be present

- if $1 \leq min \leq max \leq |H(e)|$, it is a OR feature group, and no more than the max and no less than the min child features can be present if the parent feature is present. Figure 10 shows the hyper-arc diagram for the feature model.

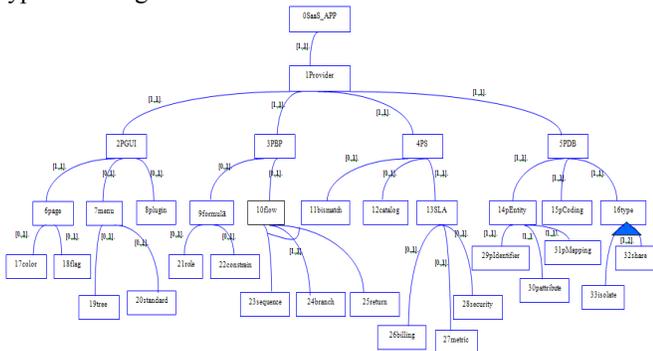

Fig 10 Hyper-arc Diagram for the Feature Model

*Formula of variation and commonality:*

We can represent the variability and commonality of each layer of SaaS application by formula 1 and 2, respectively.

$$variability = \frac{k}{2^n - 1} ............ (1)$$

$k$: is number of products

$n$: is number of all features

Variability increase the number of tenants and cost

$$commonality = \frac{sharenode}{k} ........ (2)$$

Sharednode: number of appeared nodes in all products

*Node in provider diagram as input:*

{0(0.SaaS_APP); 1(1.Provider); 2(2. PGUI); 3(3.pBP); 4(4.PS);5(5.PDB);6(6.page);7(7.menu);8(8.plugin);9(9.formula);10(10.folow);11(11.dismatch);12(12.catalog);13(13.SLA);14(14.pEntity);15(15.pCoding);16(16.type); 17(17.color);18(18.flag);19(19.tree);20(20.standard);21(21.role);22(22.constraint);23(23.sequence);24(24.branch);25(25.return);26(26.billing);27(27.metric);28(28.security);29(29.pIdentifier);30(30.pAttribute);31(31.pMapping);32(32.share); 33(33.isolate) };

## 6.2 Provider Level Hyper-arcs

We described the vertices and edges for all models. Also we showed the relationship between the vertices and determined all groups belonging to any vertices in table 1. Hyperarcs: From[mult]H{To}: as input

Table 1 Inputs of Vertexes and Relationship

| Node | Relation | Childs Group | Node | Relation | Childs Group |
|---|---|---|---|---|---|
| 0 | [1,1] | {1} | 5 | [1,1] | {14 } |
| 1 | [1,1] | {2 } | 5 | [1,1] | {15} |
| 1 | [1,1] | {3 } | 5 | [1,1] | {16 } |
| 1 | [1,1] | {4 } | 6 | [0,1] | {17,18 } |
| 1 | [1,1] | {5 } | 7 | [0,1] | {19,20} |
| 2 | [1,1] | {6 } | 9 | [0,1] | {21,22 } |
| 2 | [0,1] | {7 } | 10 | [1,1] | {23,24,25} |
| 2 | [0,1] | {8 } | 13 | [1,1] | {26} |
| 3 | [0,1] | {9} | 13 | [1,1] | {27} |
| 3 | [0,1] | {10 } | 13 | [1,1] | {28} |
| 4 | [0,1] | {11,12 } | 14 | [1,1] | {29,30,31} |
| 4 | [1,1] | {13 } | 16 | [1,1] | {32,33} |

From our description of the configuration for the provider level we have observed that there are many variations. Those will help SaaS application to give multiple choices and provide many tenants. We should input the data as the system can make self-configuration by table 2.

Table 2 SCAS Algorithm of SaaS Application Layers

| Algorithm Name: SCAS |
|---|
| Inputs: $n$ : nodes of all model, Relation: *Relationship* between nodes, Group :all item belong to any nodes. $L$: application layers(*GUI, BP, S, DB*) Outputs: All configurations for layers |
| 1 For each $l \in L$ |
| 2    // children's cardinality set is more than one |
| 3       While $H(e) > 1$ do |
| 4       // alternative constraint. |
| 5          if $min = 1$ and $max$=1 Then only one node |
| 6       // optional or mutex constraint |
| 7          if $min = 0$ and $max$= 1 Then in configuration can select or not |
| 8    //  OR constraint |
| 9          if $l \leq min \leq max \leq |H(e)|$ Then will select all or a part of nodes |
| 10          end if |
| 11          end if |
| 12          end if |
| 13   // children's cardinality is one |
| 14       While $H(e) = 1$ do |
| 15   // mandatory or require constraint |

| | |
|---|---|
| 16 | if *min* = 1 and *max*= 1 then must select in configuration |
| 17 | // optional constraint |
| 18 | if *min* = 0 and *max*= 1 then may select or not |
| 19 | end if |
| 20 | end if |
| 21 | configure(*l*) |
| 22 | // return number of SaaS Layer configurations |
| 23 | *k* = count configuration (*l*) |
| 24 | // return number of all nodes in full configuration |
| 25 | *n*=count node (*l*) |
| 26 | // calculate variability of layers components |
| 27 | $V = \frac{k}{2^n - 1}$ |
| 28 | // calculate commonality of layers components |
| 29 | sharenode=count sharenode(configure(l)). |
| 30 | $C = \frac{sharenode}{k}$ |
| 31 | end while |
| 32 | end while |
| 33 | if *H(e)* = 0 |
| 34 | invalid configuration |
| 35 | end if |
| 36 | end for each |

In our running example the system needs to reconfigure because the application exchanges from time to time and from travel agency to travel agency and from user to user. As an example, in high season booking will need different configurations to realize all travel agency requirements, which depends on the variation of customer requirements. In this time the travel agency needs to offer various options for booking like different rates. For that we can monitor the variability of every layer as we mentioned above and make decisions to best configure and realized the agency requirements. In addition we can monitor the commonality of any component in every layer to show the degree of sharing of this component in different configurations.

The algorithm can dynamically configure every layer to show all options that show the variation of the configurations from tenant to tenant. However, to calculate the variability and commonality they will be an indicator to monitor the system configuration.

## 7    Related Work

The direction of the work is for the meta-model and modeling for the evolution of SaaS application. In [12] defined the criteria for designing the process model and realized commonality and variability of modeling to maximize the reusability. Researchers in [13] analyzed tenancy history metadata from the graphic user interface (GUI), workflow, service, and data layer for dynamically adjusting template objects. In [14] provided an on-demand service-oriented model driven architecture to develop an enterprise mashup prototype as a practical case study. Authors in [15] regarded PIM can be used to generate different PSMs using transformation tools to minimize the time, cost and efforts in developing cloud SaaS and enhance the return on investment. They identified technical issues and proposed their effective solution spaces in [16]. In [17] proposed a QoS model and

MCDM (Multi Criteria Decision Making) system for SaaS ERP. They empirically examined main drivers and inhibiting factors of SaaS-adoption for different application types in [18]. In [19] studied forecasts effects expected when the SaaS model will be fully applied to the library network. And they presented functional requirements and an operation model of SaaS-based library management systems. In [20] extensible business component model named xBC is proposed for describing both the structural and behavioral properties of generic SaaS applications to minimized the amount of sources needed to be reexamined by a transformation when the source is changed. All development and evolution done by meta-model, but it is not mention how to enable model-driven development and tool support for the integration of self-management functionality into SaaS application.

## 8    Conclusions

This research is a foundation to build a new model for SaaS application. By meta-modeling it defined four layers to composite system and showed the associations and dependencies of the layer elements. We have demonstrated the relationship between three levels in our model by a workflow model as a business process layer. We observed the necessity of sharing the workflow in every level which can improve efficiency and better control service to the consumer. In our new model we could classify services of SaaS application according to three levels. We have increased the quality of the system by showing it has different level services and can serve by important ordering. From meta-model layers we have conducted the variation of the element layers and can obtain different configuration than other methods [21]. In addition we have described the self-configuration algorithm to dynamically configure SaaS components.

In future work we will see the effect of our new model to QoS in  the SaaS application.

## 9    Acknowledgement

This work has been developed with the support under the project with number: 2012AA02A604, 863 Program key projects in China: The Technology and the System Development for Smart Acquirement of Personal Healthcare Information. And so the Key Project of NSF in China: Methodology of Value-oriented Software Services: Theory, Method and Application with number: 61033005

## Appendix – XML file of feature model

```
<?xml version="1.0" encoding="UTF-8" standalone="no"?>
<featureModel chosenLayoutAlgorithm="1">
<struct>          <and          abstract="true"          mandatory="true"
name="SaaS_APP"><and mandatory="true" name="provider"><and
mandatory="true"      name="PGUI">      <and      mandatory="true"
name="page"><feature name="color"/><feature name="flag"/>
</and>          <and name="menu"><feature name="tree"/><feature
name="standard"/></and>   <feature name="plugin"/></and>   <and
mandatory="true" name="PBP">   <and name="formula">
     <feature name="role"/> <feature name="constraint"/>
     </and><or      name="flow">      <feature      mandatory="true"
name="sequence"/><feature mandatory="true" name="branch"/>
     <feature mandatory="true" name="return"/>
     </or></and><and mandatory="true" name="PS">
     <feature name="bispatch"/>
     <feature      name="catalog"/>      <and      mandatory="true"
name="SLA">   <feature name="billing"/>
     <feature      name="meteric"/>          <feature      name="security"/>
</and></and><and mandatory="true" name="PDB">
     <and mandatory="true" name="pEntity">
<feature mandatory="true" name="pIdentifier">
     <feature mandatory="true" name="pAttribute"/>
     <feature mandatory="true" name="pMapping"/>
     </and>   <feature mandatory="true" name="pCoding"/>
     <or mandatory="true" name="type">
     <feature mandatory="true" name="share"/>
     <feature mandatory="true" name="isolate"/>
     </or></and></and><and          mandatory="true"
name="tenant">   <feature mandatory="true" name="TGUI">
     <feature mandatory="true" name="TBP"/>
     <feature mandatory="true" name="TS"/>
     <feature mandatory="true" name="TDB"/>
     </and><and mandatory="true" name="user">
     <feature mandatory="true" name="UGUI"/>
     <feature mandatory="true" name="UBP"/>
     <feature mandatory="true" name="US"/>
     <and mandatory="true" name="UDB">
          <feature mandatory="true" name="uEntity">
          <feature mandatory="true" name="uIdentifier"/>
          <feature mandatory="true" name="uAttribute"/>
          <feature mandatory="true" name="uMapping"/>
          </and>
          <feature mandatory="true" name="uCoding"/>
          </and> </and></and></and> </struct>
<constraints/><comments/>
<featureOrder userDefined="false"/></featureModel>
```